\documentstyle[11pt,paspconf]{article}

\def\ltsim{\raisebox{-.5ex}{$\;\stackrel{<}{\sim}\;$}}
\def\gtsim{\raisebox{-.5ex}{$\;\stackrel{>}{\sim}\;$}}

\def  \La          {\ifmmode {\rm Ly}\alpha \else Ly$\alpha$\fi}
\def  \Ha          {\ifmmode {\rm H}\alpha \else H$\alpha$\fi}
\def  \Hb          {\ifmmode {\rm H}\beta \else H$\beta$\fi}
\def  \CIIIb       {\ifmmode {\rm C}\,{\sc iii]} \else C\,{\sc iii]}\fi}
\def  \CIV         {\ifmmode {\rm C}\,{\sc iv} \else C\,{\sc iv}\fi}
\def  \SiIV         {\ifmmode {\rm Si}\,{\sc iv} \else Si\,{\sc iv}\fi}
\def  \MgII         {\ifmmode {\rm Mg}\,{\sc ii} \else Mg\,{\sc ii}\fi}
\def  \HeII         {\ifmmode {\rm He}\,{\sc ii} \else He\,{\sc ii}\fi}
\def  \NV         {\ifmmode {\rm N}\,{\sc v} \else N\,{\sc v}\fi}

\begin{document}

\noindent{\small Invited review to appear in {\it ``Quasars and Cosmology'',
18-22 May 1998, La Serena Chile.}
A.S.P. Conference Series~1999.
eds. G.Ferland, J.Baldwin. (San Francisco: ASP)}

\title{Echo Mapping of AGN Emission Regions}
\author{Keith Horne}
\affil{University of St.Andrews, Physics \& Astronomy,
North Haugh, St.Andrews KY16~9LS, Scotland, UK}

\begin{abstract}
Echo mapping exploits light travel time delays,
revealed by multi-wavelength variability studies,
to map the geometry, kinematics, and physical conditions
of reprocessing sites in photo-ionized gas flows.
In AGNs, the ultra-violet to near infra-red light arises in part
from reprocessing of EUV and X-ray light from 
a compact and erratically variable source in the nucleus.
The observed time delays, 0.1-2 days for the continuum,
1-100 days for the broad emission lines,
probe regions only micro-arcseconds away from the nucleus.
The continuum time delays map the temperature-radius
profiles of the AGN accretion discs.
The emission-line delays reveal radially stratified ionization zones,
identify the nature of the gas motions, 
and estimate the masses of the central black holes.
By using light travel time to measure the sizes
of AGN accretion discs and photo-ionized zones,
echo mapping offers two independent ways to
measure redshift-independent distances to AGNs.
\end{abstract}


\section{Introduction}
\label{sec:introduction}

Following this brief introduction,
section \ref{sec:methods} reviews in general terms
the techniques used for echo mapping of active galaxies.
Sections \ref{sec:geometry}, \ref{sec:kinematics}, and \ref{sec:conditions} 
then discuss the use of echo mapping experiments to probe
the geometry, the kinematics, and the physical conditions
in Seyfert 1 galaxies.
Section \ref{sec:h0} considers two direct methods based on
echo mapping to determine redshift-independent distances,
and hence cosmological parameters $H_0$ and $q_0$.

\subsection{The Black Hole Accretion Disc Model for AGNs}

The standard model of an Active Galactic Nucleus (AGN)
envisions a supermassive ($10^{6-9}M_\odot$) black hole
in the core of a galaxy.
The black hole is fed gas from an accretion disc.
Around the disc is a geometrically thick torus of
dust and molecular gas.
The disc and torus receive gas from the galaxy's interstellar medium,
and from various processes (tides, bow shocks, winds, irradiation)
that strip material from the stars that pass through this region.

The ultra-violet and optical spectra of quasars,
and their lower-luminosity cousins in the nuclei Seyfert galaxies,
have two types of emission lines:
narrow forbidden and permitted emission lines ($v \ltsim 1000~$km~s$^{-1}$),
and broad permitted emission lines ($v \gtsim 10,000~$km~s$^{-1}$).
The narrow lines are constant; the continuum and the broad lines vary.
The narrow lines arise from lower-density gas at larger 
distances from the nucleus.

Seyfert nuclei come in two types thought to represent
different viewing angles.
Seyfert 1 spectra exhibit both broad and narrow emission lines,
while in Seyfert 2 spectra the broad lines are absent
or visible only in a faint linearly polarized component of the spectrum.
The current consensus is that the thick dusty torus blocks
our view of the BLR when viewed at $i\gtsim60^\circ$, but
some BLR light scatters toward us after rising far enough above
the plane to clear the torus.

The Narrow Line Region (NLR) in nearby Seyfert galaxies
is resolved by {\it HST} imaging studies 
with resolutions of $\sim 0.1$~arcseconds.
The NLR typically has a clumpy bi-conical morphology,
aligned with the axis of radio jets, and suggesting 
broadly-collimated ionization cones emerging from the nucleus,
perhaps collimated by the dusty torus.
The BLR and continuum production regions are unresolved by {\it HST}.
Echo mapping experiments use light travel time delays
revealed by variability on 1-100 day timescales
to probe these regions on $\sim 10^{-6}$ arcsecond scales.

\subsection{Photo-Ionization Models}

Just how large is the BLR?
If the emission lines are powered by photo-ionization,
then we may employ a model to estimate the size of the
photo-ionized region.
Photo-ionization models such as CLOUDY
(Ferland, et al.~1998)
consider the energy and ionization balance inside a 1-dimensional
gas cloud parameterized by hydrogen number density $n_{\sc H}$,
column density $N_{\sc H}$, and distance $R$ from the source of
ionizing radiation $L_\lambda$.
The calculated emission-line spectrum emerging from such a
gas cloud depends primarily on the ionization parameter,
\begin{equation}
	U = \frac{\rm ionizing\ photons}{\rm target\ atoms}
	= \frac{Q}{4\pi R^2 c n_{\sc H}} ,
\end{equation}
where the luminosity of hydrogen-ionizing photons is
\begin{equation}
	Q = \int_{0}^{\lambda_0} \frac{\lambda L_\lambda d\lambda}{h c} ,
\end{equation}
with $L_\lambda$ the luminosity spectrum emitted by the nucleus.
Re-arranging this equation yields
\begin{equation}
	R = \left( \frac{Q}{4\pi U c n_{\sc H} }\right)^{1/2} .
\end{equation}
By comparing the flux ratios observed for the broad emission lines
in the spectra of AGNs 
against those predicted by the single-cloud photo-ionization models,
typical parameters in the photo-ionized gas are found to be
$U \sim 10^{-2}$ and $n_{\sc H} \sim 10^{8-10}$cm$^{-3}$.
With these conditions, the light travel time
across the radius of the ionized zone is
\begin{equation}
\label{eqn:blrsize}
	\frac{R}{c} \sim 200 {\rm d}
	\left( \frac{Q}{10^{54}{\rm s}^{-1} } \right)^{1/2}
	\left( \frac{U}{10^{-2} } \right)^{-1/2} 
	\left( \frac{n_{\sc H}}{10^{9}{\rm cm}^{-3} } \right)^{-1/2} ,
\end{equation}
where $Q\sim 10^{54}$s$^{-1}(H_0/100)^{-2}$ is appropriate for
the Seyfert 1 nucleus of NGC~5548.
On the basis of such predictions, astronomers searched
for reverberation effects on timescales of months
before eventually realizing that reverberation was occurring
on much shorter timescales.

\section{Echo Mapping Methods}
\label{sec:methods}

\begin{figure}
\plotfiddle{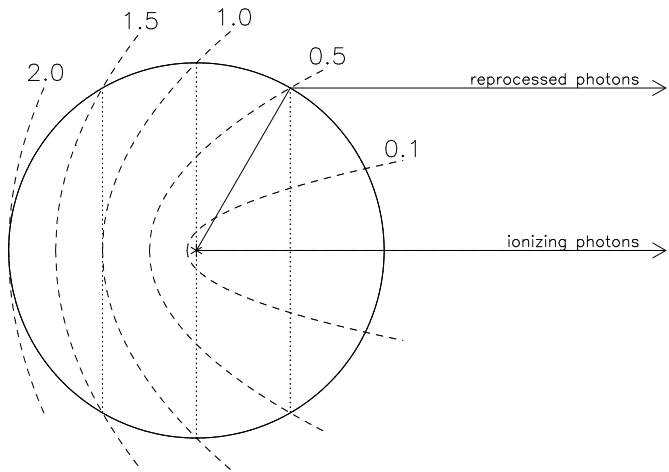}{5cm}{-90}{45}{45}{-200}{240}
\caption{ 
Ionizing photons from a compact source are reprocessed
by gas clouds in a thin spherical shell.
A distant observer sees the reprocessed photons arrive
with a time delay ranging from 0, for the near edge of the shell,
to $2R/c$, for the far edge.
The iso-delay paraboloids slice the shell into zones with
areas proportional to the range of delays.
}
\label{fig:shell}
\end{figure}

\subsection{Reverberation}

A compact variable source of ionizing radiation
launches spherical waves of heating or cooling that
expand at the speed of light, triggering
changes in the radiation emitted by surrounding gas
clouds.
The light travel time from nucleus to reprocessing site to observer
is longer than that for the direct path from nucleus to observer.
Thus a distant observer sees the reprocessed radiation arrive
with a time delay reflecting its position within the source.
The time delay for a gas cloud located at a distance $R$
from the nucleus is
\begin{equation}
	\tau = \frac{R}{c} ( 1 + \cos{\theta} )\ ,
\end{equation}
where the angle $\theta$ is measured from 0 for a cloud on the
far side of the nucleus to $180^\circ$
on the line of sight between the nucleus and the observer.
The delay is 0 for gas on the line of sight between us and
the nucleus, and $2R/c$ for gas directly behind the nucleus
(Fig.~\ref{fig:shell}).
The iso-delay contours are concentric paraboloids wrapped
around the line of sight.
Such time delays are the basis of echo mapping techniques.

The validity of this basic picture is well supported by the results of
intensive campaigns designed to monitor the variable optical,
ultra-violet, and X-ray spectra of Seyfert 1 galaxies.
Notable among these are the AGN~Watch campaigns
(\verb+http://www.astronomy.ohio-state.edu/~agnwatch/+).
In Seyfert 1 galaxies, 
the variations observed in continuum light are practically simultaneous
at all wavelengths throughout the ultra-violet and optical,
but the corresponding emission line variations
exhibit time delays of 1 to 100 days,
probing the photo-ionized gas 1 to 100 light days from the nucleus.
This corresponds to micro-arcsecond scales in nearby Seyfert galaxies.

\subsection{Linear and Non-Linear Reprocessing Models}

The ionizing radiation that drives the line emission
includes X-ray and EUV light that cannot be directly observed.
However, quite similar variations are seen in continuum
light curves at ultra-violet and optical wavelengths.
These observable continuum light curves provide suitable surrogates
for the unobservable ionizing radiation.
The reprocessing time (hours) is small compared
with light travel time (days) in AGNs.

In the simplest reprocessing model,
the line emission $L(t)$ is taken to be proportional to
the continuum emission $C(t-\tau)$, with a time delay
$\tau$ arising from light travel time and from local
reprocessing time, thus
\begin{equation}
        L(t) = \Psi C(t-\tau)\ .
\end{equation}
Since the reprocessing sites span a range of time delays,
the line light curve $L(t)$ is a sum of many time-delayed copies
of the continuum light curve $C(t)$,
\begin{equation}
	L(t) = \int_0^\infty \Psi(\tau)\ C(t-\tau)\ d\tau .
\end{equation}
This convolution integral introduces $\Psi(\tau)$,
the ``transfer function'' or ``convolution kernel'' or ``delay map'',
to describe the strength of the reprocessed light
that arrives with various time delays.

Since light travels at a constant speed, the surfaces of constant
time delay are ellipsoids with one focus at
the nucleus and the other at the observer.
Near the nucleus, these ellipsoids are effectively paraboloids
(Fig.~\ref{fig:shell}).
$\Psi(\tau)$ is in effect a 1-dimensional map
that slices up the emission-line gas,
revealing how much gas is present between each of the
iso-delay paraboloids.

The aim of echo mapping is to recover $\Psi(\tau)$ from 
measurements of $L(t)$ and $C(t)$ made at specific times $t_i$.
To fit such observations,
the linear model above is too simple in at least two respects.
The first problem is additional sources of light contributing
to the observed continuum and emission-line fluxes.
Examples are background starlight, and narrow emission lines.
When these sources do not vary on human timescales,
they simply add constants to $L(t)$ and $C(t)$,
\begin{equation}
\label{eqn:linearized}
	L(t) = \bar{L} + \int_0^\infty 
		\Psi(\tau)\ 
		\left[ C(t-\tau) - \bar{C} \right] 
		d\tau .
\end{equation}

\begin{figure}
\plotfiddle{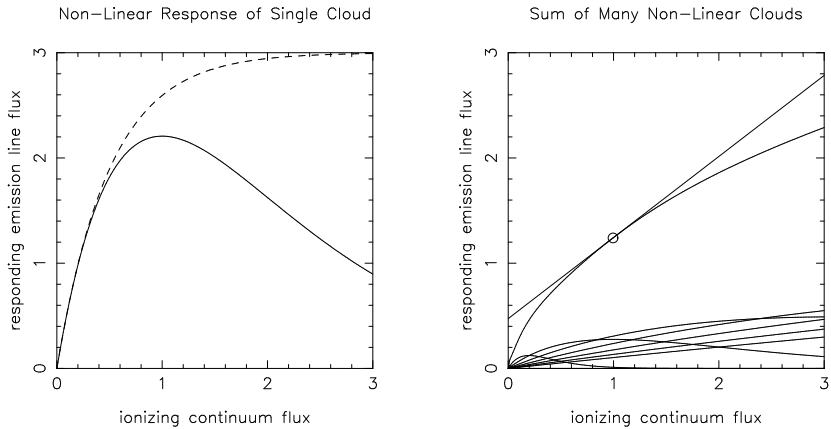}{5cm}{-90}{45}{45}{-180}{230}
\caption{ 
The emission-line response from gas clouds
exposed to different levels of photo-ionizing continuum radiation.
Left: The initially linear response saturates
and may even decrease as the cloud becomes more fully ionized.
Right: When gas clouds are present at many radii,
the ionization zone can simply expand, resulting in
an ensemble response that increases monotonically
and is less non-linear than that from a single cloud.
The tangent line is then a useful approximation.
}
\label{fig:nonlin}
\end{figure}

A second problem with the linear reprocessing model is that
the reprocessed emission is more generally a non-linear function
of the ionizing radiation.
For example, when a cloud is only partially ionized, the line
emission increases with the ionizing radiation,
but as the cloud becomes fully ionized
the line emission may saturate or even decrease
with further increases in ionizing radiation.
For single clouds we therefore expect the response function $L(C)$
to be highly non-linear (Fig.~\ref{fig:nonlin}).
The marginal response $\partial L / \partial C$
is generally less than the mean response $L/C$,
and may become negative when an increase in ionizing radiation
reduces the line emission.

The total response is of course a sum of responses
from many different gas clouds, those closer to the nucleus
being more fully ionized than those farther away.
If clouds are present at many radii, so that zones of constant
ionization parameter can simply expand, then the
effect of averaging over many clouds is to produce a total response
that is monotonically increasing and less strongly non-linear.
Such a response may be adequately approximated by a tangent line,
\begin{equation}
	L(C) = \bar{L} + \frac{\partial L}{\partial C}
			\left( C - \bar{C} \right)\ ,
\end{equation}
for ionizing radiation changes in some range above and below a mean level
(Fig.~\ref{fig:nonlin}).
Observations support the use of this approximation --
plots of observed line fluxes against continuum fluxes
show roughly linear relationships,
with $L/C > \partial L / \partial C$ so that 
extrapolation to zero continuum flux leaves a positive
residual line flux.

For these reasons, it is usually appropriate in echo mapping
to adopt the linearized reprocessing model,
equation~\ref{eqn:linearized}, with
the ``background'' fluxes, $\bar{L}$ for the line
and $\bar{C}$ for the continuum, set somewhere in the
range of values spanned by the observations.
In this model, the delay map $\Psi(\tau)$ senses each gas cloud
in proportion to its marginal response to a change in ionizing radiation.
The roughly linear responses from partially-ionized clouds
are fully registered, 
while the saturated or diminished responses of more fully ionized clouds
have a reduced effect.

\subsection{Cross-Correlation Analyses}

\begin{figure}
\plotfiddle{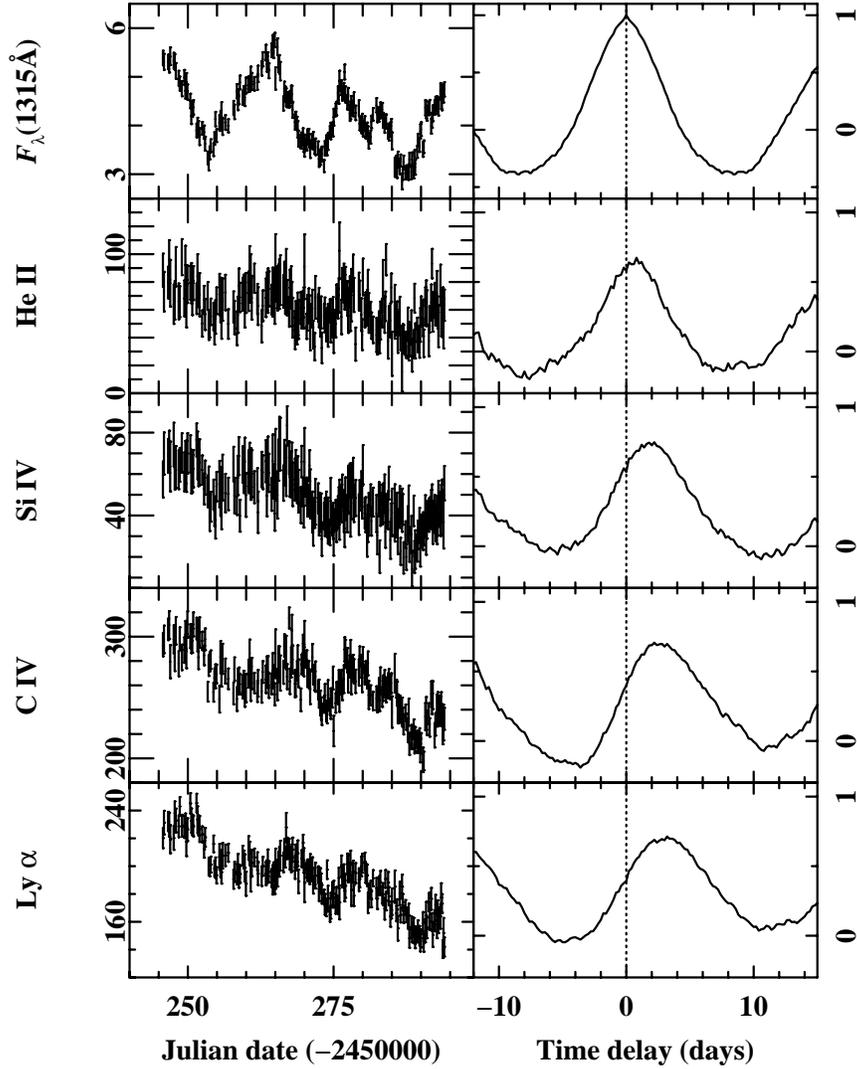}{15cm}{0}{60}{60}{-200}{-20}
\caption{ 
The left-hand columns show light curves of NGC~7469 obtained
with {\it IUE} during an intensive AGN~Watch monitoring campaign
during the summer of 1996.
The right-hand column shows the result of cross-correlating
the light curve immediately to the left with the
1315~\AA\ light curve at the top of the left column;
the panel at the top of the right column thus shows
the 1315~\AA\ continuum auto-correlation function.
Data from Wanders, et~al.~(1997).
}
\label{fig:ccf}
\end{figure}

With reasonably complete light curves, it is usually obvious that
the highs and lows in the line light curves occur later
than those in the continuum.
To quantify this time delay, or lag, a common practice is
to cross-correlate the line and continuum light curves.
The cross-correlation function (CCF) is
\begin{equation}
	L \star C (\tau) = \int L(t)\ C(t-\tau)\ dt .
\end{equation}
Several methods have been developed and refined
to compute CCFs from noisy measurements available only
at discrete unevenly-spaced times,
either by interpolating the data
(Gaskell \& Sparke~1986, White \& Peterson~1994)
or binning the CCF (Edelson \& Krolik 1988, Alexander 1997).
The resulting CCFs generally have a peak
shifted away from zero in a direction indicating that changes in
the emission lines lag behind those in the continuum
(e.g.\ Fig.~\ref{fig:ccf}).
The CCF lag -- $\tau_{\sc CCF}$ --
at which $L \star C$ has its maximum value, serves to
quantify roughly the size of the emission-line region.

Since a range of time delays is present, described by $\Psi(\tau)$,
what delay is measured by the cross-correlation peak?
Since $L(t)$ is itself a convolution between $C(t)$ and $\Psi(\tau)$,
and since convolution is a linear operation,
we have
\begin{equation}
	L \star C = ( \Psi \star C ) \star C = \Psi \star ( C \star C ).
\end{equation}
The CCF is therefore a convolution
of the delay map with the continuum auto-correlation function (ACF),
\begin{equation}
	C \star C(\tau) = \int C(t)\ C(t-\tau)\ dt .
\end{equation}
The ACF is symmetric in $\tau$, and thus always
has a peak at $\tau=0$.
With rapid continuum variations, the ACF is sharp,
and the CCF peak should be close to
the strongest peak of $\Psi(\tau)$.
This tends to favour short delays from the inner regions of the BLR
(Perez, et~al. 1992a).
When continuum variations are slow, however,
the ACF is broad, smearing out
sharp peaks in $\Psi(\tau)$, and shifting $\tau_{\sc CCF}$
toward the centroid of $\Psi(\tau)$.
Thus the cross-correlation peak depends not only on the
delay structure in $\Psi(\tau)$, but also on the
character of the continuum variations.
Different observing campaigns may therefore yield different lags
even when the underlying delay map is the same.

\subsection{Three Echo Mapping Methods}

More refined echo mapping analyses aim to recover
the delay map $\Psi(\tau)$ rather than just a characteristic time lag.
These echo mapping methods generally require more complete data
than the cross-correlation analyses.
Three practical methods have been developed.

The {\bf Regularized Linear Inversion} method (RLI)
(Vio, et~al.~1994, Krolik \& Done~1995)
notes that the convolution integral,
\begin{equation}
	L(t) = \int C(t-\tau)\ \Psi(\tau)\ d\tau\ ,
\end{equation}
becomes a matrix equation,
\begin{equation}
	L(t_i) = \sum_j C(t_i-\tau_j)\ \Psi(\tau_j)\ d\tau\ ,
\end{equation}
when the times $t_i$ are evenly spaced.
If the times are not evenly spaced, interpolate.
If the matrix $C_{ij} = C(t_i-\tau_j)$ can be inverted,
solving the matrix equation yields
\begin{equation}
	\Psi(\tau_k) = \sum_k C^{-1}_{ki}  L(t_i) / d\tau .
\end{equation}
If the continuum variations are unsuitable,
the matrix has small eigenvalues and the inversion is unstable,
strongly amplifying noise in the measurements $L(t_i)$ and $C(t_i)$.
This problem is treated by altering the matrix to reduce the
influence of small eigenvalues, 
thereby reducing noise but blurring the delay map.

The related method of {\bf Subtractive Optimally-Localized Averages} (SOLA)
(Pijpers \& Wanders 1994) aims to estimate the delay map
as weighted averages of the emission line measurements,
\begin{equation}
	\hat{\Psi}(\tau) = \int K(\tau,t)\ L(t)\ dt.
\end{equation}
Since $L= C\star \Psi$, we may write this as
\begin{equation}
	\hat{\Psi}(\tau) = \int K(\tau,t) \int C(t-s)\ \Psi(s)\ ds\ dt\ .
\end{equation}
The estimate $\hat{\Psi}$ is therefore a blurred version of
the true delay map $\Psi$.
With suitable continuum variations,
the weights $K(\tau,t)$ can be chosen to make the blur kernel
\begin{equation}
	b(\tau,s) = \int K(\tau,t)\ C(t-s)\ dt
\end{equation}
resemble a narrow Gaussian of width $\Delta$ centred
at $s = \tau$.
The parameter $\Delta$ then controls the trade-off between noise
and resolution in reconstructing $\Psi(\tau)$.
These direct inversion methods work best when the observed
light curves detect continuum variations on a wide range of timescales.

The {\bf Maximum Entropy Method} (MEM)
(Horne, Welsh \& Peterson 1991, Horne~1994)
is a very general fitting method
allowing the use of any linear
or non-linear reverberation model.
As an extension of maximum likelihood techniques,
MEM employs a ``badness of fit'' statistic, $\chi^2$,
to judge whether the echo model being considered
achieves a satisfactory fit to the data.
The requirement $\chi^2/N \sim 1 \pm \sqrt{2/N}$,
where $N$ is the number of continuum and line measurements,
ensures that model fits as well as is
warranted by the error bars on the data points,
without over-fitting to noise.
MEM fits are required also to maximize the
`entropy', which is designed to measure the `simplicity' of the map.
For positive maps $p_i>0$, the entropy
\begin{equation}
	S = \sum_i p_i - q_i - p_i \ln{p_i/q_i}
\end{equation}
is maximized when $p_i = q_i$.
MEM thus steers map $p_i$ toward default values $q_i$.
With the default map set to
\begin{equation}
	q_i = \left( p_{i-1} p_{i+1} \right)^{1/2},
\end{equation}
the entropy steers each pixel toward its neighbors,
and MEM then finds the `smoothest' positive map that fits the data.
For further technical details see Horne (1994).

\begin{figure}
\plotfiddle{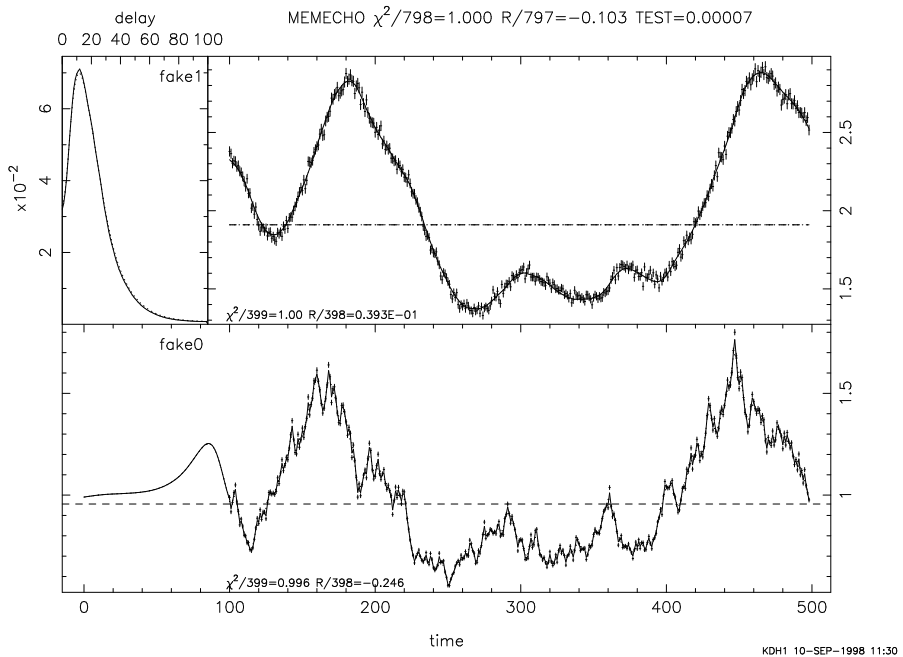}{6cm}{-90}{45}{45}{-180}{252}
\caption{ Simulation test of the maximum entropy method
showing the recovery of a delay map (top left) from
data points sampling an erratically varying continuum light
curve (bottom) and the corresponding delay-smeared
emission-line light curve (top right).
The reconstructed delay map closely resembles
the true map $\Psi(\tau) \propto \tau e^{-\tau}$.
}
\label{fig:faketest}
\end{figure}

Fig.~\ref{fig:faketest} illustrates recovery of a delay map
from a MEM fit to fake light curves.
The lower panel shows an erratically varying continuum light curve.
The line light curve in the upper panel is smoother and has
time-delayed peaks.
A smooth continuum light curve $C(t)$ threads through the data points,
and extrapolates to earlier times.
Convolving this light curve with the delay map $\Psi(\tau)$, shown in the
upper left panel, gives the line light curve $L(t)$, 
fitting the data points in the upper right panel.
Dashed lines give the backgrounds $\bar{L}$ and $\bar{C}$.
The MEM fit with $\chi^2/N=1$ adjusts $C(t)$, $\Psi(\tau)$, and $\bar{L}$
to fit the data points while keeping $C(t)$ and $\Psi(\tau)$ as
smooth as possible.

In this simulation test, the true transfer function,
$\Psi(\tau) \propto \tau e^{-\tau}$, is accurately recovered
from the data.
The fake dataset represents the type of data that could be obtained
with daily sampling over a baseline of 1 year.
While this is rather better than has been achieved so far,
future experiments (Kronos, Robonet) specifically designed for
long-term monitoring will make this simulation more relevant.

\section{Mapping the Geometry of Emission-Line Regions}
\label{sec:geometry}

\subsection{ Spherical Shells }

\begin{figure}
\plotfiddle{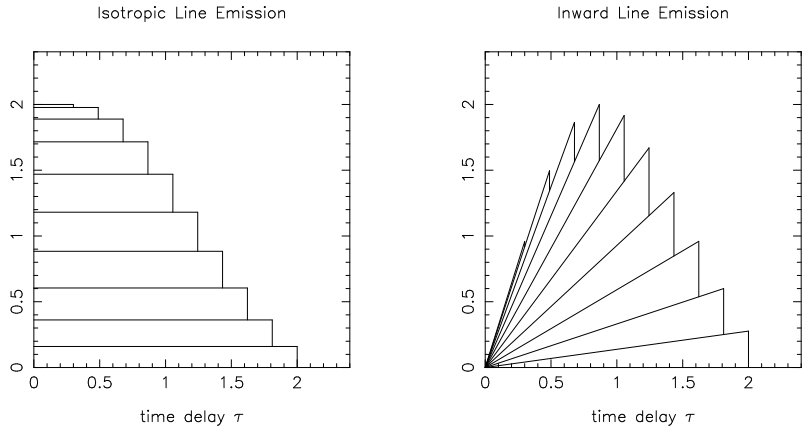}{6cm}{-90}{45}{45}{-180}{230}
\caption{ 
Delay maps for spherical geometry.
Left: When the line emission is isotropic, 
the contribution from each spherical shell is 
constant from 0 to $2R/c$.
The total delay map, summing many spherical shells, must
decrease monotonically.
Right: When the line emission is directed inward, toward the nucleus,
the response at small delays is reduced, so that each shell's
contribution increases with $\tau$ between 0 and $2R/c$.
The total delay map then rises to a peak away from zero.
}
\label{fig:psi}
\end{figure}

Having shown that we can recover the delay map $\Psi(\tau)$,
given suitable data, how may we interpret it?
This is relatively straightforward if the geometry is 
spherically symmetric.

Consider a thin spherical shell that is irradated by a brief flash
of ionizing radiation from a source located at the shell's centre.
The flash reaches every point on the shell after a time $R/c$.
Since recombination times are short,
each point responds by emitting a brief flash of recombination radiation.
A distant observer sees first the flash of ionizing radiation,
and then the response of reprocessed light from the shell
arriving with a range of time delays.
The time delay is 0 for the near edge of the shell,
and $2R/c$ on the far edge of the shell.
The iso-delay paraboloids 
slice up the spherical shell
into zones with areas proportional to the range of time delay
(Fig.~\ref{fig:shell}).
The response from the spherical shell is therefore
a boxcar function,
constant between the delay limits 0 and $2R/c$
(Fig.~\ref{fig:psi}).

Any spherically symmetric geometry is just
a nested set of concentric spherical shells.
The delay map for a spherical geometry is therefore
a sum of boxcar functions (Fig.~\ref{fig:psi}).
Since all the boxcars begin at delay 0,
the delay map must peak at delay 0,
and decrease monotonically thereafter.
The contribution from the shell of radius $R$ 
may be identified from the slope $\partial \Psi / \partial \tau $
evaluated at the appropriate time delay $\tau = 2R/c$.
(The case of anisotropic emission is discussed
in Section~\ref{sec:aniso} below.)

\subsection{The BLR is Smaller than Expected}

\begin{figure}
\plotfiddle{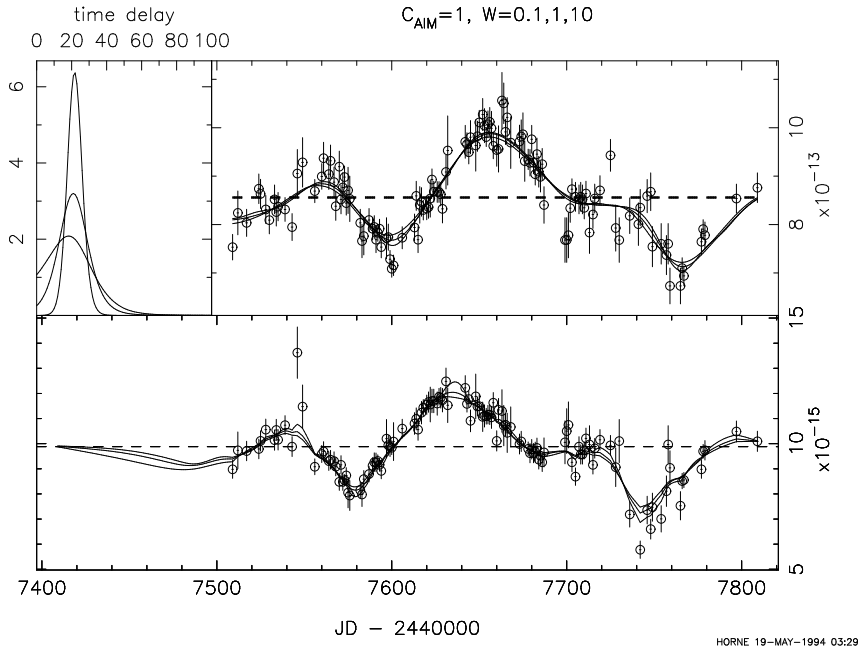}{7cm}{-90}{45}{45}{-180}{252}
\caption{ 
Echo maps of \Hb\ emission in NGC~5548 found by a maximum entropy
fit to data points from a 9-month AGN~Watch monitoring campaign
during 1989.
The optical continuum light curve (lower panel) is convolved with
the delay map (top left) to produce the \Hb\ emission line
light curve (top right).
Horizontal dashed lines give the mean line and continuum fluxes.
Three fits are shown to indicate likely uncertainties.
Data from Horne, Welsh \& Peterson~(1991).
}
\label{fig:hbmap}
\end{figure}

Fig.~\ref{fig:hbmap} shows a maximum entropy fit
of the linearized echo model of Eqn.~\ref{eqn:linearized}
to \Hb\ and optical continuum
light curves of NGC~5548 (Horne, Welsh \& Peterson 1991).
Subtracting the continuum background $\bar{C}$,
convolving with the delay map $\Psi(\tau)$,
and adding the line background $\bar{L}$,
gives the \Hb\ light curve.
The three fits shown, all with $\chi^2/N=1$,
indicate the likely range of uncertainty
due to the trade-off between the `stiffness' of
$\Psi(\tau)$ and $C(t)$.

The \Hb\ map has a single peak at a delay of 20 days,
and declines to low values by 40 days.
This suggests that the size of the \Hb\ emission-line
region is 10-20 light days.
This is 10 to 20 times smaller than the 200 light day size
estimated in Eqn.~\ref{eqn:blrsize}
on the basis of single-cloud photo-ionization models.
With clouds this close to the nucleus, the
ionization parameter $U$ would be higher unless
gas densities are increased by a factor of 100,
to $n_{\sc H} \sim 10^{11}$cm$^{-3}$.

\subsection{Anisotropic Emission}
\label{sec:aniso}

The \Hb\ response in NGC~5548 is smaller at delay 0 than at 20 days.
This lack of a prompt response conflicts with the
monotonically decreasing delay map we expect for a
spherical geometry.
One interpretation is that there is a deficit of gas
on the line of sight to the nucleus.

However, more likely this is a signature of anisotropic
emission of the \Hb\ photons arising from optically
thick gas clouds that are photo-ionized only on their
inward faces (Ferland, et~al. 1992, O'Brien, et~al. 1994).
The preference for inward emission of \Hb\ photons
reduces the prompt response by making
the reprocessed light from clouds on the far side
stronger than that from clouds on the near side of the nucleus.
With inward anisotropy, the impulse response of a spherical shell
is a wedge, increasing with $\tau$,
rather than a boxcar (Fig.~\ref{fig:psi}).
The sum of wedges has a peak away from zero.

\subsection{Stratified Temperature and Ionization}

Ultra-violet spectra from {\it IUE} and {\it HST} provide shorter-wavelength
continuum light curves, and light curves for a variety of
high and low ionization emission lines.
These are useful probes of the temperature and
ionization structure of the reprocessing gas.

The continuum light curves display very similar structure
at all optical and ultraviolet wavelengths.
The continuum lightcurve at the shortest ultra-violet wavelength
observed is normally used as the light curve against which to
study time delays at other continuum wavelengths and in
the emission lines.
In most cases studied to date the optical and ultra-violet continuum
light curves exhibit practically simultaneous variations
(e.g. Peterson, et~al. 1998), implying
that the continuum production regions are unresolved,
smaller than a few light days.

In one case, NGC~7469, delays of 0.1 to 1.7 days are detected
between the 1350\AA\ and 7500\AA\ continuum light curves
based on 40 days of continuous {\it IUE} and optical monitoring 
(Wanders et~al. 1997, Collier et~al. 1998).
The delay increases with wavelength as $\tau \propto \lambda^{4/3}$,
suggesting that the temperature decreases as $T \propto R^{-3/4}$.

A systematic pattern is generally seen in the emission-line time delays.
High-ionization lines (\NV, \HeII) exhibit the shortest delays,
while lower-ionization lines (\CIV,\La,\Hb) have longer delays.
This was established in the first AGN~Watch campaign,
by cross-correlation analysis (Clavel et~al. 1991)
and by maximum entropy fits (Krolik, et~al. 1991),
using {\it IUE} and optical light curves of NGC~5548
sampled at 4 day intervals for 240 days.
The effect is now seen in many objects (Table \ref{tab:lags}).
This pattern is consistent with photo-ionization models,
in which higher ionization zones occur closer to the nucleus.

\begin{table}

\begin{center}
\caption{\bf AGN~Watch Monitoring Campaigns
\label{tab:lags}
}
\begin{tabular}{crccccl}
\hline
target & year & \multicolumn{4}{c}{$\tau_{\sc CCF}$ (days)}& references
\\ & & \HeII\ & \La\ & \CIV\ & \Hb\ &
\\ \hline \hline 
   NGC~5548 & 1989 & 7 & 12 & 12 & 20 & 1, 2, 3
\\ NGC~3783 & 1992 & 0 &  4 &  4 &  8 & 4, 5
\\ NGC~5548 & 1993 & 2 &  8 &  8 & 14 & 6
\\ Fairall~9 & 1994 & 4 & 17 & -- & 23  & 7, 8
\\ 3C~390.3 & 1995 & 10 & 50 & 50 & 20 & 11, 12
\\ NGC~7469 & 1996 & 1 & 3 & 3 & 6 & 9, 10 
\\ \hline
\end{tabular}
\end{center}

\noindent References:
	1. Clavel, et~al.~1991;
	2. Peterson, et~al.~1991;
	3. Dietrich, et~al.~1993;
	6. Korista, et~al.~1995;
	4. Reichert, et~al.~1994;
	5. Stirpe, et~al.~1994;
	7. Rodriguez-Pascual, et~al.~1997;
	8. Santos-Lleo, et~al.~1997;
	9. O'Brien, et~al.~1998;
	10. Dietrich, et~al.~1998;
	11. Wanders, et~al.~1997;
	12. Collier, et~al.~1998.
\end{table}

Another interesting result is that
the time delay for \CIV\ emission
is generally smaller than that for \CIIIb\ emission.
In early work, this line ratio was the basis for
estimating the gas density in the BLR to be $n_{\sc H} \sim 10^{8-10}$cm$^{-3}$.
However, this practice is now considered dubious
because their different time delays indicate that
these lines arise in different regions,
The \CIV\ line arises from gas clouds closer to the nucleus,
and a higher density, $n_{\sc H} \sim 10^{11}$cm$^{-3}$,
is therefore needed to maintain the ionization parameter.

\subsection{Ionized Zones Expanding with Luminosity}

\begin{figure}
\plotfiddle{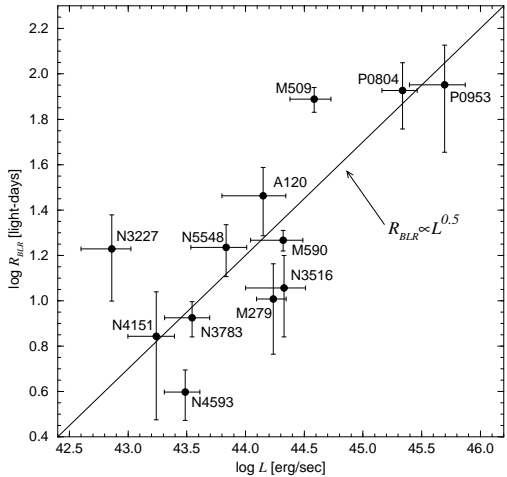}{6cm}{00}{45}{45}{-130}{-20}
\caption{ 
The radii of ionized zones emitting broad \Hb\ emission lines
for AGNs with different ionizing continuum luminosities.
The radii are estimated from time delays that are
found by cross-correlating \Hb\ and optical continuum light curves.
The diagonal line with $R \propto L^{1/2}$
corresponds to a constant ionization parameter.
Figure from Kaspi, et~al. (1996).
}
\label{fig:kaspi}
\end{figure}

The ionization zones should be larger in higher luminosity objects.
This prediction has been tested using
echo mapping studies of a dozen active galaxies,
including two low-luminosity quasars, spanning three decades in luminosity
(Fig.~\ref{fig:kaspi}, Kaspi, et~al. 1996).
The lag for each object is determined by cross-correlating
\Hb\ and optical continuum light curves.
A log-log plot of the time delay vs luminosity
shows a standard deviation of about 0.3 dex about a correlation
of the form
\begin{equation}
	\tau_{\sc \Hb} \sim 17{\rm d}\ L_{44}^{1/2} ,
\end{equation}
where $L_{44}$ is the hydrogen-ionizing luminosity
in units of $10^{44}$erg~s$^{-1}$.
For comparison,
if gas clouds with similar densities are present
over a wide range of radius,  
then zones of constant ionization parameter
should increase with $R \propto L^{1/2}$.

In the above correlation, derived by comparing different objects,
the observed scatter may be due in part to differences
in inclination, black hole mass, and accretion rate.
The ionized zone in each object is expected expected to change size
if the ionizing luminosity changes by a substantial factor.
There is some evidence for this, e.g. from 8 years of monitoring
NGC~5548 (Peterson, et~al. 1999).
To map this effect, the echo model may be generalized to allow the
delay map to change with the driving continuum light curve,
\begin{equation}
	L(t) = \int_0^\infty C(t-\tau)\ \Psi(\tau,C(t-\tau))\ d\tau\ .
\end{equation}

\section{Mapping the Kinematics of Emission-Line Regions}
\label{sec:kinematics}

\begin{figure}
\plotfiddle{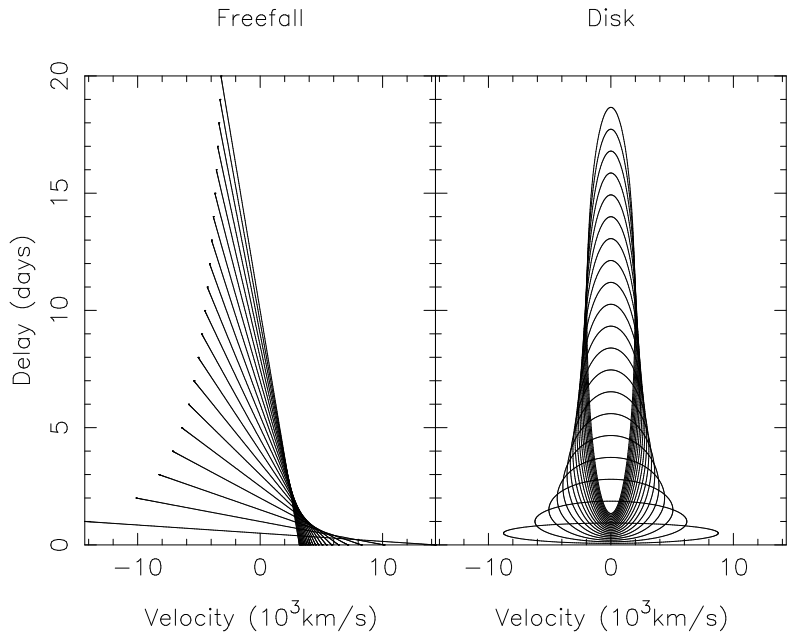}{7cm}{-90}{45}{45}{-180}{252}
\caption{ 
Velocity--delay maps for spherical freefall
and Keplerian disc kinematics.
Spherical shells map to diagonal lines,
and disc annuli map to ellipses.
The central mass is $10^{7}$M$_\odot$ in both cases.
The disc
 inclination angle is $60^\circ$.
}
\label{fig:vtmaps}
\end{figure}

Combining time delays from variability with Doppler shifts
from emission-line profiles
yields a velocity-delay map $\Psi(v,\tau)$ to
probe the kinematics of the flow.
Fig.~\ref{fig:vtmaps} illustrates this capability by sketching
the formation of velocity-delay maps for
spherical free-fall into a point-mass potential,
and for a Keplerian disc
(Welsh \& Horne 1991, Perez, et~al. 1992b).
The dramatically different appearance of the two flows
highlights the power of velocity-delay maps for kinematic
diagnosis.

A radial flow, inward or outward, produces a strong red--blue asymmetry
in the velocity-delay map, while an azimuthal flow does not.
Line emission from gas flowing in toward  the nucleus
is redshifted ($v>0$) on the near side ($\tau$ small)
and blueshifted ($v<0$) on the far side ($\tau$ large).
The signature of inflowing gas is therefore small time delays
on the red side and large time delays on the blue side.
Outflowing gas produces just the opposite red--blue asymmetry.
Gas circulating around the nucleus has the same time delay
on the red and blue side, producing a symmetric velocity-delay map.

Any spherically-symmetric flow is
a nested set of thin spherical shells.
The time delay $\tau$ and Doppler shift $v$ are
\begin{equation}
\begin{array}{ccc}
       \tau = \frac{R}{c} \left( 1 + \cos{\theta} \right) ,
& ~~~~ &
       v = v_R \cos{\theta} ,
\end{array}
\end{equation}
with $R$ the shell radius,
$\theta$ the angle from the back edge,
and $v_R$ the outflow velocity.
The linear dependence of both $\tau$ and $v$ on $\cos{\theta}$
implies that each shell maps into a diagonal line in the velocity-delay
plane, as shown in Fig.~\ref{fig:vtmaps}.

The circulating disc
 flow is a set of
concentric co-planar cylindrical annuli.
The time delay and Doppler shift are
\begin{equation}
\begin{array}{ccc}
       \tau = \frac{R}{c} \left( 1 + \sin{i} \sin{\phi} \right) ,
& ~~~~ &
       v = v_\phi \sin{i} \cos{\phi} ,
\end{array}
\end{equation}
with $R$ and $i$ the radius and inclination of the annulus,
$\phi$ the azimuth,
and $v_\phi$ the azimuthal velocity.
Each annulus maps to an ellipse on the velocity-delay plane,
as in Fig.~\ref{fig:vtmaps}.
Inner annuli map to ``squashed'' ellipses
(large $\pm v$, small $\tau$),
while outer annuli map to ``stretched'' ellipses
(small $\pm v$, large $\tau$).

While $\Psi(v,\tau)$ is a distorted 2-dimensional
projection of the 6-dimensional phase space,
it can reveal important aspects of the flow, particularly if
the velocity field is ordered and has some degree of symmetry.
The envelope of the velocity-delay map may reveal
the presence of virial motions $v^2 \propto G M / c \tau $,
with $M$ the mass of the central object.
A red/blue asymmetry, or lack thereof, gauges the relative
importance of radial and azimuthal motions.
Disordered velocity fields with a range of velocities
at each position smear the map in the $v$ direction.
The far side (large $\tau$) may be enhanced by
anisotropic emission from optically thick clouds
radiating their lines inward toward the nucleus
(Ferland, et~al. 1992, O'Brien, et~al. 1994).

To derive Doppler-delay maps from the data,
we simply slice the observed line profile into wavelength bins,
and recover the delay map at each wavelength.
This is a simple extension of the echo model
used to fit continuum and emission-line light curves.
At each wavelength $\lambda$ and time $t$ the emission-line flux
\begin{equation}
	L(\lambda,t) = \bar{L}(\lambda) +
	\int_0^\infty \left[ C(t-\tau) - \bar{C} \right]\
	\Psi(\lambda,\tau)\  d\tau
\end{equation}
is obtained by summing time-delayed responses
to the continuum light curve and adding 
a time-independent background spectrum $\bar{L}(\lambda)$.

We observe $C(t)$ and $L(\lambda,t)$ at some set of times $t_i$.
We fix $\bar{C}$, e.g. close to the mean of the observed values.
We then adjust the continuum light curve $C(t)$,
the background spectrum $\bar{L}(\lambda)$,
and the Doppler-delay map $\Psi(\lambda,\tau)$,
in order to fit the observations.
The fitting procedure maximizes the entropy
to find the ``simplest'' map(s) that fit the
observations with $\chi^2/N = 1$.

\begin{figure}
\plotfiddle{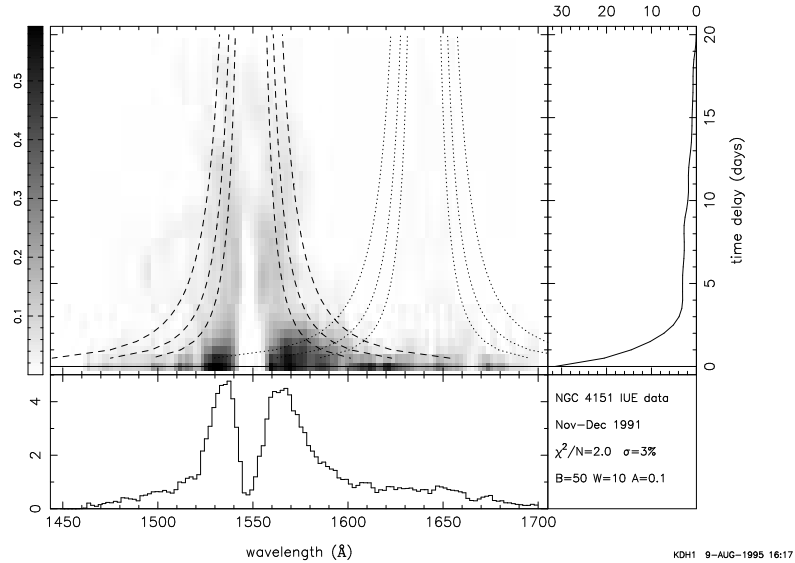}{6cm}{-90}{40}{40}{-180}{240}
\caption{ {\sc memecho } velocity-delay map of \CIV\ 1550 emission
(and superimposed \HeII\ 1640 emission) in NGC~4151.
Dashed curves give escape velocity for masses
0.5, 1.0, and $2.0\times10^7$M$_\odot$.
}
\label{fig:4151}
\end{figure}

Fig.~\ref{fig:4151} exhibits the velocity-delay map for \CIV\ 1550
and \HeII\ 1640 emission as reconstructed from
44 {\it IUE} spectra of NGC~4151
covering 22 epochs during 1991 Nov 9 -- Dec 15
(Ulrich \& Horne 1996).
While spanning only 36 days, this campaign recorded
favorable continuum variations, including a bumpy exponential decline
followed by a rapid rise, which were sufficient to support echo mapping on
delays from 0 to 20 days.

Disregard the strong \CIV\ absorption feature, which
obliterates the delay structure at small velocities.
The wide range of velocities at small delays
and smaller range at larger delays suggests virial motions.
The dashed curves in Fig.~\ref{fig:4151} give escape
velocity envelopes
$v = \sqrt{2GM/c\tau}$ for masses 0.5, 1.0, and $2.0 \times
10^{7}$~M$_\odot$.
A mass of order $10^{7}$~M$_\odot$
may be concentrated within 1~light-day of the nucleus.

The approximate red--blue symmetry of the \CIV\ map
rules out purely radial inflow or outflow kinematics.
The somewhat stronger \CIV\ response
on the red side at small delays
and on the blue side at larger delays
suggests a gas component with freefall kinematics.
However, if the \CIV\ emission arises from the irradiated faces of
optically-thick clouds, those on the far side of the nucleus
will be brighter, and the redward asymmetry can then be interpreted
as an outflow combined with the inward \CIV\ anisotropy.

Velocity-delay maps have so far been constructed
only for \CIV\ emission in
NGC~5548 (Wanders, et~al. 1995, Done \& Krolik 1996)
and
NGC~4151 (Ulrich \& Horne 1996).
Both systems show the same trend of velocity dispersion
decreasing with delay,
and red response stronger at small delays.
As more maps are constructed, we will learn whether these
are general characteristics of Seyfert broad-line regions.

\section{Mapping Physical Conditions in Emission-Line Regions}
\label{sec:conditions}

\subsection{Quasar Tomography}

The rich information coded in high quality time-resolved spectrophotometry
of a reverberating emission-line spectrum
can be extracted only by fitting observations in far
greater detail than has previously been attempted.
Here we extend previous echo mapping methods to map simultaneously
the geometry, kinematics, and physical conditions
characterizing the population of photo-ionized gas clouds.
To do this we fit simultaneously the complete
reverberating spectrum, including the reverberations observed
in the fluxes and profiles of numerous emission lines.
These fits incorporate explicitly the predictions of 
a photo-ionization model such as CLOUDY,
thus accounting for the non-linear and anisotropic responses
of emission-line clouds to changes in the ionizing radiation.

We characterize a gas cloud by its density $n_{\sc H}$,
column density $N_{\sc H}$, distance from the nucleus $R$,
azimuth $\theta$, and Doppler shift $v$.
The entire population of gas clouds is then
described by a 5-dimensional map
$\Psi( R, \theta, n_{\sc H}, N_{\sc H}, v )$.
We omit the cloud's position angle $\phi$ around the line of sight,
and the perpendicular velocity components,
because the data provide no useful constraint on these.

Assuming that the shape of the ionizing spectrum is known,
and that the time-dependent ionizing photon luminosity is $Q(t)$,
then the ionization parameter obtaining at time $t$ is
\begin{equation}
	U(t) = \frac {Q(t-\tau)} {4 \pi R^2 n_{\sc H} c} .
\end{equation}
The reprocessing efficiency $\epsilon_L(U,n_{\sc H},N_{\sc H},\theta)$
for emission line $L$ depends in a unique way on $U$,
$n_{\sc H}$, and $N_{\sc H}$.
The reprocessing efficiencies
are evaluated with a photo-ionization code, e.g.\ CLOUDY,
for chosen element abundances.
To save computer time, these are pre-calculated 
on a suitable grid of cloud parameters,
e.g. equally spaced in $\log{U}$, $\log{n_{\sc H}}$, and $\log{N_{\sc H}}$
(Korista, et~al. 1997).
Results required for any cloud parameters are
subsequently found rapidly by interpolation in the grid.

Because clouds are irradiated on their inward faces,
the reprocessed line emission is generally anisotropic.
We allow for this by letting $\epsilon_L$ depend on $\theta$.
CLOUDY computes the emission emerging inward, $\theta=0$,
and outward, $\theta=90^\circ$.
For intermediate angles we interpolate linearly in $\cos{\theta}$.

The observed spectrum is a sum of three components:
direct light from the nucleus,
reprocessed light from the surrounding gas clouds,
and background light, -- e.g. from stars,
\begin{equation}
	f_\nu(\lambda,t) =
	f^D_\nu(\lambda,t) + f^R_\nu(\lambda,t) + f^B_\nu(\lambda)\ .
\end{equation}
The direct light is
\begin{equation}
	f^D_\nu(\lambda,t) =
	\frac { Q(t) S_\nu(\lambda) } { 4 \pi D^2 }
\end{equation}
where $D$ is the distance and the spectral shape is
\begin{equation}
        S_\nu(\lambda) = \frac{ L_\nu(\lambda,t) }{ Q(t) } .
\end{equation}

The reprocessed light is a sum over numerous emission lines,
where $\lambda_L$ and $\epsilon_L$ are the
rest wavelength and the reprocessing efficiency for line $L$.
The gas clouds are distributed over a 3-dimensional volume.
At each location the cloud population has a distribution
over density $n_{\sc H}$, column density $N_{\sc H}$, and Doppler shift $v$.
The reprocessed light arising from such a configuration is
\begin{equation}
\begin{array}{rl}
f^R_\nu(\lambda,t)
	= & \int 2\pi R dR\ d\mu\ dn_{\sc H}\ dN_{\sc H}\ dv\
	\Psi(R,\theta,n_{\sc H},N_{\sc H},v)\

\\ & 	\frac{Q(t-\tau)}{4 \pi D^2}\
	\sum_L \epsilon_L(U,n_{\sc H},N_{\sc H},\theta)\ 
	G_\nu(\lambda-(1+v/c)\lambda_L)\
\\ &	\delta\left( R - \left( \frac{ Q(t-\tau) } { 4 \pi U n_{\sc H} }
			\right)^{1/2} \right)\
	\delta\left( \mu - \frac{\tau c}{R} + 1 \right)\ .
\end{array}
\end{equation}
Here $\mu = \cos{\theta}$,
$G_\nu$ is a Gaussian profile to apply the appropriate Doppler shift,
and the Dirac $\delta$ functions ensure that the
appropriate time delay and ionization parameter are used
at each reprocessing site.

To fit the above model to observations of $f_\nu(\lambda,t)$,
we adjust $D$, $Q(t)$, $f^B_\nu(\lambda)$,
and $\Psi(R,\theta,n_{\sc H},N_{\sc H},v)$.
The 5-dimensional cloud map $\Psi$ can have loads of pixels,
$\sim 10^{5-6}$.
The constraints available from reverberating emission-line spectra
will not be sufficient to fully determine the cloud map.
However, once again MEM may be employed
to locate the ``smoothest'' and ``most symmetric'' 
cloud maps that fits the data.
Computers are now fast enough to support
this type of detailed modelling and mapping of AGN emission regions.

\subsection{Mapping a Spherical Shell from Simulated Data}

\begin{figure}
\plotfiddle{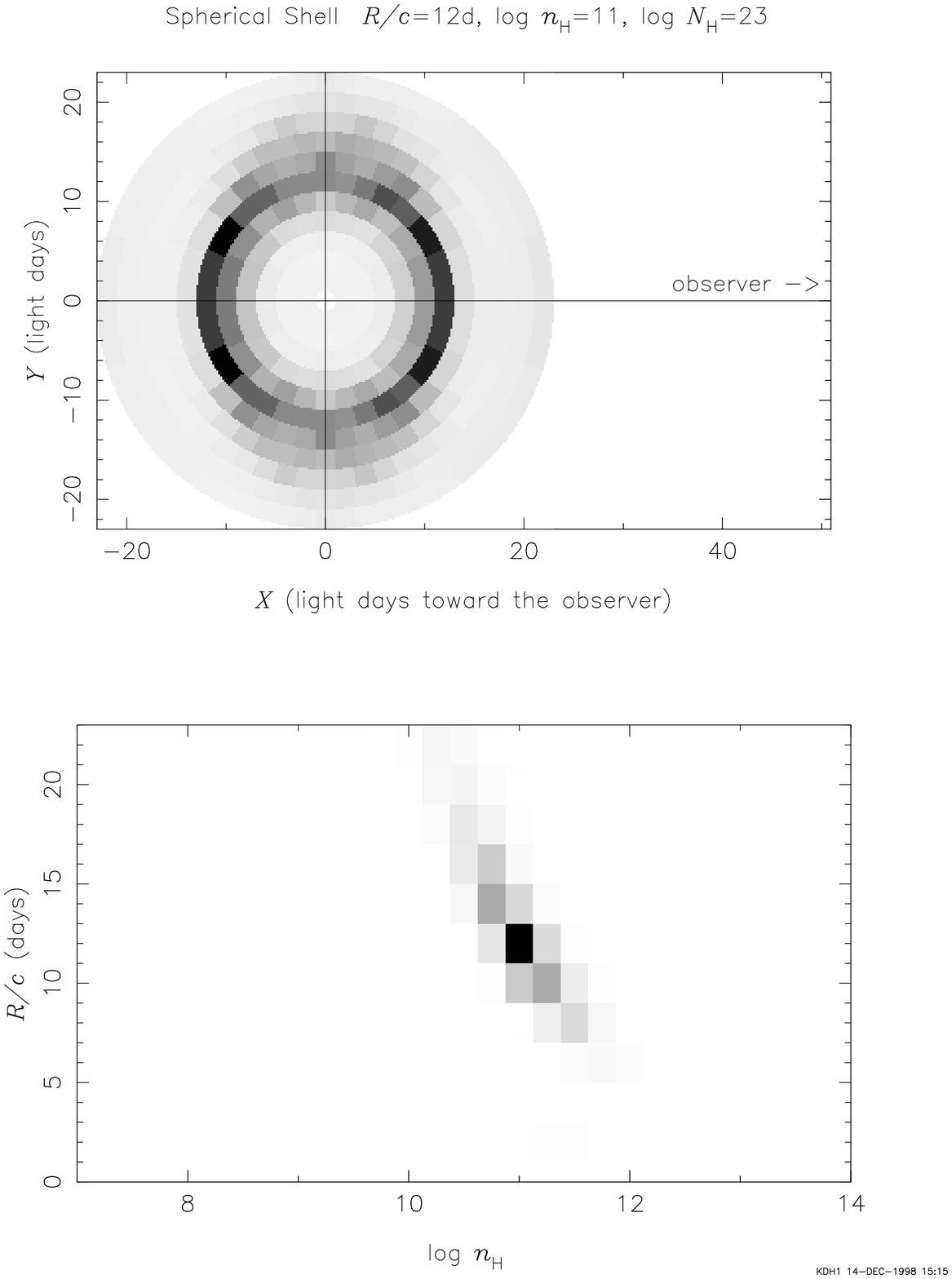}{6.5cm}{0}{35}{35}{-205}{-28}
\plotfiddle{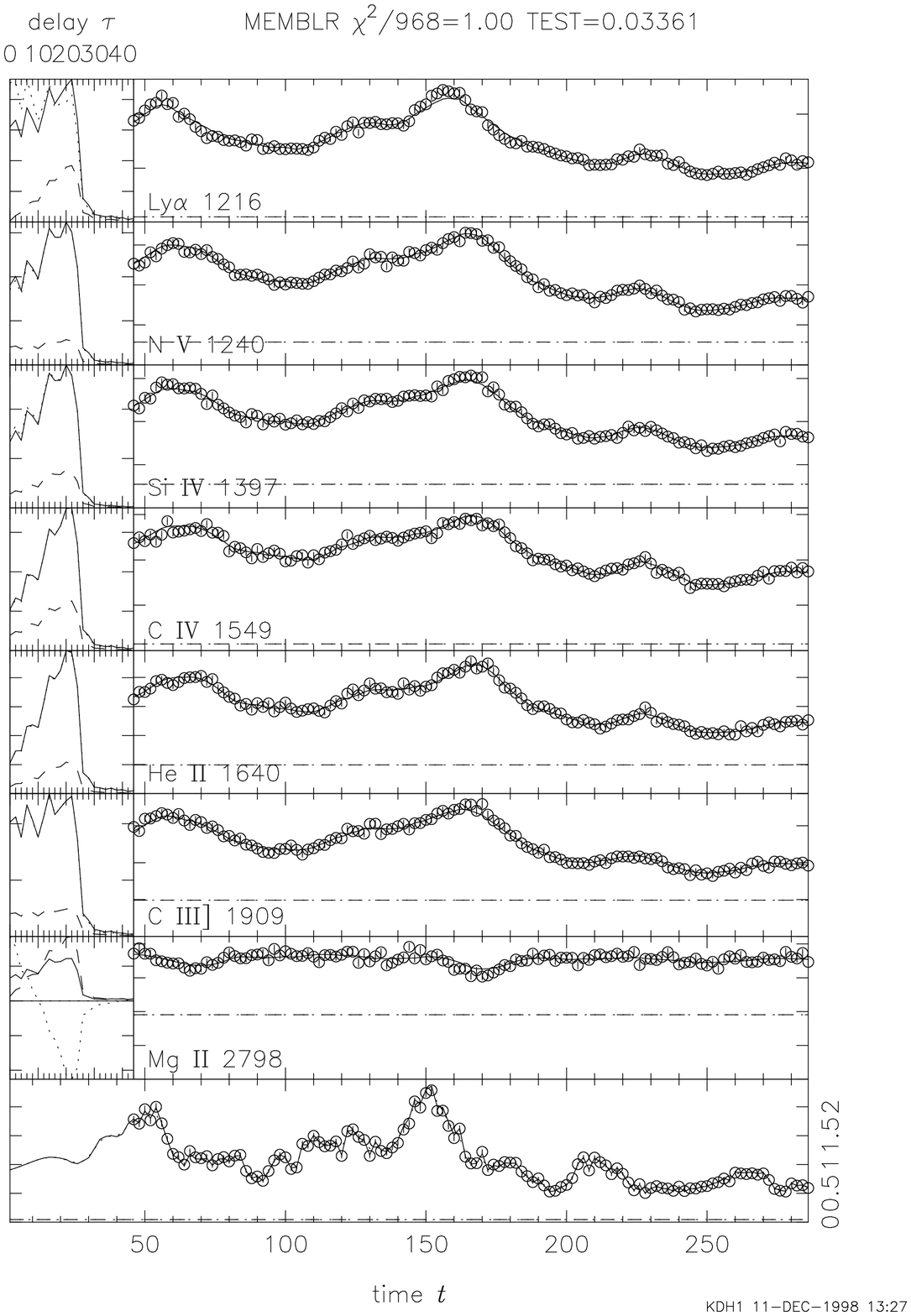}{0cm}{0}{35}{35}{-25}{-5}
\caption{ 
Reconstructed map $\Psi(R,\theta,n_{\sc H})$
of a thin spherical shell (left) from a maximum entropy
fit to 7 ultra-violet emission-line light curves (right).
The light curve fits achieve $\chi^2/N=1$.
The shell radius $R/c = 12$d and density $\log{n_{\sc H}}=11$
are correctly recovered.
For further details, see text.
}
\label{fig:shellfit}
\end{figure}

Fig.~\ref{fig:shellfit} shows a simulation test designed to determine
how well the geometry and physical conditions may be recoverable
from emission-line reverberations.
The adopted BLR model places clouds with density
$n_{\sc H}=10^{11}$cm$^{-3}$ and column density $N_{\sc H}=10^{23}$cm$^{-2}$
on a thin spherical shell of radius $R/c=12$d.
Synthetic light curves are computed showing reverberations in
7 ultra-violet emission lines, using CLOUDY to calculate the
appropriate anisotropic and non-linear emissivities.
The synthetic light curves are sampled at 121 epochs spaced by 2 days,
and noise is added to simulate observational errors.

The MEM fit does not assume a spherical shell geometry,
but rather it considers every possible cloud map $\Psi(R,\theta,n_{\sc H})$,
and tries to find the simplest such map that fits the 
emission line light curves.
The MEM fit adjusts 4147 pixels in the cloud
map $\Psi(R,\theta,n_{\sc H})$,
143 points in the continuum light curve $C(t)$, 
7 emission-line background fluxes $\bar{L}(\lambda)$, and
1 continuum flux $\bar{C}$.
Note that the fit assumes the correct column density and distance.
The fit to $N=968$ data points achieves $\chi^2/N=1$.
The entropy steers each pixel in the map
toward its nearest neighbors, thus encouraging smooth maps, 
and toward the pixel with the opposite sign of $\cos{\theta}$,
to encourage front-back symmetry.

On the left-hand side of Fig.~\ref{fig:shellfit}, we see that the
recovered geometry displays the appearance of a hollow shell
with the correct radius, and that the density-radius projection
of the map has a peak at the correct density.
The shell spreads in radius by a few light days,
with lower densities at larger radii to
maintain the same ionization parameter.

On the right-hand side of Fig.~\ref{fig:shellfit}, we see that
the 8 light curves are well reproduced by the fit.
The highs and lows are a bit more extreme in the data --
a common characteristic of regularized fits.
To the left of each emission-line light curve, three
delay maps are shown corresponding to the maximum brightness
(solid curve), minimum brightness (dashed), 
and the difference (dotted).
All the lines exhibit an inward anisotropy except \CIIIb.
All the lines have positive linear responses except \MgII,
which has a positive response on the near side of the shell
and a negative response on the far side.

This simulation test has used reverberation effects in
emission-line light curves to map the geometry and density structure
in the photo-ionized emission line zone.
The 2-dimensional map clearly reveals the correct hollow shell geometry,
and moreover the correct density is also recovered.
These promising results suggest that there are good prospects for probing
the geometry and physical conditions in real AGNs.

\section{Two New Direct Routes to $H_0$}
\label{sec:h0}

Because the primary aim of this meeting is to discuss ways in which
quasars may be useful as standard candles for cosmology,
I will conclude with a brief discussion of two methods,
both relatively new, which may allow the determination of
redshift-independent distances to active galaxies.
Both methods employ echo mapping time delays to measure the physical
size of something.
Emission line time delays measure the sizes of
regions of ionized gas surrounding the nucleus.
Continuum time delays measure the sizes of reprocessing
zones on the reverberating surface of an irradiated accretion 
disc.

\subsection{$H_0$ from Sizes of Photo-Ionization Zones}
\label{sec:h0_lines}

The ionizing photon luminosity $Q$ can be estimated
from an observed flux $F$ as
\begin{equation}
	Q = 4\pi D^2 S F\ ,
\end{equation}
where $D$ is the distance and $S$ is a ``spectral shape'' factor
that converts the observed flux $F$ to an ionizing photon flux
\begin{equation}
	S F = \int_0^{\lambda_0}
	\frac{ \lambda f_\lambda d\lambda }{ h c }\ .
\end{equation}
If the ionizing radiation is not isotropic, the flux $F$
in our direction differs from that seen by the photo-ionized gas clouds.
That uncertainty may be absorbed into the factor $S$.

From the definition of the ionization parameter $U$, 
the radius of the photo-ionized zone is given by
\begin{equation}
	R = \left( \frac{Q}{4\pi U n_{\sc H} c } \right)^{1/2}
	= D \left( \frac{ S F }{ U n_{\sc H} c } \right)^{1/2}\ .
\end{equation}
Reverberation results allow us to measure $R = \tau c$
from a time delay $\tau$.
We therefore find that the distance is
\begin{equation}
	D = \tau \left( \frac { U n_{\sc H} c^3 } { S F } \right)^{1/2} .
\end{equation}
Since $\tau$ is measured by reverberation, while
$U$, $n_{\sc H}$, $S$, and $F$ are found from the emission-line spectrum,
we have a means of estimating the distance
without using the redshift.
For redshift $z << 1 $, the Hubble constant is then
\begin{equation}
	H_0 = \frac{cz}{D} 
	= \frac{z}{\tau} \left( \frac {S F} { U n_{\sc H} c } \right)^{1/2} .
\end{equation}

The above results apply only for the over-simplified case
of photo-ionized gas characterized by single values of
the ionization parameter $U$ and density $n_{\sc H}$.
This is intended only as an illustrative sketch of the method.
In a more realistic model, the gas clouds have not single values but rather
a cloud distribution function $\Psi(R,\theta,n_{\sc H},N_{\sc H})$.
We discussed this type of fitting in Section~\ref{sec:conditions}.

\subsection{$H_0$ from Reverberating Accretion Discs}
\label{sec:h0_continuum}

\noindent{\bf Steady-State Accretion Discs.}
The effective temperature on the surface of a steady-state accretion 
disc decreases with radius as
\begin{equation}
	T = \left( \frac{3 G M \dot{M}}{ 8 \pi \sigma R^3 }
\right)^{1/4} ,
\end{equation}
where $M$ is the black hole mass and $\dot{M}$ is the accretion rate.
If the disc surface radiates as a blackbody, 
the spectrum arising from the disc is
obtained by summing blackbody spectra 
weighted by the projected areas of the annuli,
\begin{equation}
f_\nu = \int B_\nu(T) \frac{ 2\pi R dR \cos{i} }{ D^2 } 
	= \left( \frac{ 1200 G^2 h }{ \pi^9 c^2} \right)^{1/3}
		I\
		\left( \frac { \cos{i} } {D^2} \right)
		\left( M \dot{M} \right)^{2/3} 
		\nu^{1/3} .
\end{equation}
Here $D$ is the distance, $i$ is the inclination of the disc
axis relative to the line of sight, and
$I = \int_0^\infty x^{5/3}/(e^x-1) \sim 1.932$.

The optical/ultra-violet spectra of AGNs have ``Big Blue Bump''
components that are attributed to thermal emission from
accretion discs.
Observed spectra are generally redder than the
characteristic $f_\nu \propto \nu^{1/3}$ spectrum predicted by disc
theory (e.g.\ Francis et~al. 1991).
The spectral signature of an accretion disc is therefore not
very convincingly demonstrated.
However, observed spectra are contaminated e.g.\
by starlight from the host galaxy (e.g.\ Welsh et~al. 1999).
Taking difference spectra cancels out the contamination and
measures the variable component of the light.
The difference spectra are usually bluer and can be in satisfactory
agreement with the predicted spectrum for the change 
in brightness of an irradiated accretion disc
(Collier et~al. 1999).

\noindent{\bf Echo Mapping Accretion Disc Temperature Profiles.}
In order to map the temperature profiles of accretion discs,
we need some way to measure the temperature, and some way
to measure the radius at which that temperature applies.
We use a time delay to measure the radius,
and a wavelength to measure the temperature.

The disc surface is irradiated by
the erratically variable source located near its centre,
launching heating and cooling waves that propogate
at the speed of light outward from the centre of the disc.
This effectively reprocesses the hard X-ray and EUV
photons that irradiate the disc
into softer ultra-violet, optical, and infra-red photons.
A distant observer will note that the reprocessed light
arrives with a time delay
\begin{equation}
	\tau = \frac{R}{c} \left( 1 + \sin{i} \cos{\theta} \right)\ .
\end{equation}
The heating wave requires a time $R/c$ to reach radius $R$,
and the reprocessing site at radius $R$ and azimuth $\theta$
is at a distance from the observer that is larger
by $R\sin{i}\cos{\theta}$
relative to the centre of the disc.

Note that for the annulus at radius $R$,
the mean time delay, averaged around the annulus,
is always $R/c$ regardless of the inclination,
but the range of time delays depends on the inclination.
If the disc is face on ($i=0$) then all azimuths have the
same time delay.
If the disc is edge on ($i=90^\circ$)
then the far edge of the disc at $\theta=0$
has the maximum time delay $\tau = 2R/c$,
and the near edge at $\theta=180^\circ$ 
has a time delay of zero.

We use the blackbody spectrum to associate each
wavelength with a temperature.
The Planck spectrum for a blackbody temperature $T$,
\begin{equation}
	B_\nu(T) = \frac{2 h c}
		{\lambda^3 \left( e^X - 1 \right) }\ ,
\end{equation}
peaks at the dimensionless frequency $X = hc / \lambda k T \sim 2.8$.
When irradiation increases $T$ slightly, 
the change in the spectrum, proportional to
$\partial B_\nu/\partial T$, peaks near $X \sim 3.8$.

To map the temperature profile of an irradiated accretion disc,
we measure the time delay at different wavelengths.
When we observe a change in the spectrum at wavelength $\lambda$,
we are measuring the reprocessed light from disc annuli
where temperatures are
\begin{equation}
	T \sim \frac{hc}{\lambda k X}
\end{equation}
with $X \sim 4$.
Ultra-violet, optical, and near infra-red light curves give
time delays for hot gas with $T\sim 10^5$K,
warm gas with $T\sim 10^4$K,
and cold gas with $T\sim 10^3$K, respectively.
If the temperature decreases with radius, the time delay
should increase with wavelength.

For the temperature profile of the steady-state disc,
the time delay is
\begin{equation}
\tau = \left( \frac{ 45 G }{ 16 \pi^6 c^5 h } \right)^{1/3}\
	\left( M\dot{M} \right)^{1/3}\
	\left( X \lambda \right)^{4/3}\ .
\end{equation}
This $\tau \propto \lambda^{4/3}$ prediction has recently been
verified in the case of the Seyfert 1 galaxy NGC~7469
(Collier, et~al. 1998).
The observed $\tau(\lambda)$ therefore allows a measurement
of $M\dot{M}$,
\begin{equation}
	M\dot{M} =
	\left( \frac { 16 \pi^6 h c^5 } { 45 G } \right)\
	\left( X \lambda \right)^{-4}\ \tau^{3}\ .
\end{equation}
Substituting for $M\dot{M}$ in the expression for $f_\nu$,
we find the distance,
\begin{equation}
D = \left( \frac{16 \pi h c^3 }{3} \right)^{1/2}\
	I^{1/3}\
	\left(\frac{\cos{i}}{f_\nu}\right)^{1/2}\
	\left( X \lambda \right)^{-4/3}\
	\tau\ .
\end{equation}
Note that this distance is determined independently of the
redshift of the AGN.
Finally, for redshift $z << 1 $, the Hubble constant is
\begin{equation}
H_0 = \frac{cz}{D} = \left( \frac {3} {16 \pi h c} \right)^{1/2}\
	I^{-1/3}
	\left( \frac {f_\nu} {\cos{i}} \right)^{1/2}\
	\left( X \lambda \right)^{4/3}\
	\left( \frac{z}{ \tau } \right)\ .
\end{equation}

The above is of course only an outline description of the method.
In practice one fits a model of reverberations in
an irradiated disc to observed light curves.
This new method is based on fairly straightforward physics --
light travel time delays to measure radius,
and blackbody spectra to associate a temperature with each wavelength.
The results for NGC~7496 give
$H_0 \sqrt{\cos{i}/0.7} = 42 \pm 9 $~km~s$^{-1}$Mpc$^{-1}$
(Collier, et~al. 1999).

The unknown disc inclination may not be too serious a problem
because at high inclinations the dusty torus should obscure the
broad emission line region.
Seyfert 1 galaxies are therefore expected to 
have $0.7 \ltsim \sqrt{\cos{i}} < 1$.
The $\sqrt{\cos{i}}$ uncertainty may be reduced further
by applying the method to a sample of Seyfert 1 galaxies
with random inclinations in the above range.
The use of blackbody spectra may be questioned, but can be checked
by seeing if the same distance is found at different wavelengths.
This test holds up for NGC~7469 (Collier, et~al. 1999).
If the method can be shown to work for a larger sample of
objects, it may serve to calibrate AGNs as standard candles
for cosmology.

\acknowledgments

Thanks to Brad Peterson and Shai Kaspi for providing figures
\ref{fig:ccf} and \ref{fig:kaspi} respectively.

\end{document}